\documentclass[aip,jcp,reprint]{revtex4-1} 
\usepackage{graphicx}

\usepackage{amsmath}    
\usepackage{graphicx}   
\usepackage{color}
\usepackage{ulem}
\usepackage{caption}
\usepackage{subcaption}

\usepackage{array}
\usepackage{booktabs}
\usepackage{multirow}

\usepackage{booktabs}

\begin{document}
\title{How wet should be the reaction coordinate for ligand unbinding?}
\author{Pratyush Tiwary}
\email{pt2399@columbia.edu} \affiliation{Department of Chemistry,
  Columbia University, New York 10027, USA.}
	
	\author{B. J. Berne}
	 \email{pt2399@columbia.edu}   
	 \affiliation{Department of Chemistry, Columbia University, New York 10027, USA.}

	\date{\today}
	
	\begin{abstract}
We use a recently proposed method called Spectral Gap Optimization of
Order Parameters (SGOOP) (Tiwary and Berne, Proc. Natl. Acad. Sci
2016 \textbf{113}, 2839 (2016)), to determine an optimal
 1-dimensional reaction coordinate (RC) for the
  unbinding of a bucky-ball from a pocket in explicit water. This RC
is estimated as a linear combination of the multiple available order
parameters that collectively can be used to distinguish the various stable states
relevant for unbinding. We pay special attention
to determining and quantifying the degree to which water
  molecules should be included in the RC. Using
SGOOP with under-sampled biased simulations, we predict that water
plays a distinct role in the reaction coordinate for unbinding in the
case when the ligand is sterically constrained to move along an axis
of symmetry. This prediction is validated through extensive
calculations of the unbinding times through metadynamics, and by
comparison through detailed balance with unbiased molecular dynamics
estimate of the binding time. However when the steric constraint is
removed, we find that the role of water in the reaction coordinate
diminishes. Here instead SGOOP identifies a good one-dimensional
RC involving various motional degrees of freedom.
\end{abstract}

	\maketitle
	\section{Introduction}
The unbinding of ligand-substrate systems is a problem of great
theoretical and practical relevance.  To take an example from the
biological sciences, there is now an emerging view that the
pharmacological efficacy of a drug depends not just on its
thermodynamic affinity for the host protein, but also, and perhaps
even more so, on when and how it unbinds from the
protein.\cite{copeland2006drug,copeland2015drug} While a variety of
experimental techniques can provide unbinding rate constants, gleaning
a clear molecular scale understanding from such experiments into the
dynamics of unbinding is difficult, and at best indirect. This makes
it in principle very attractive to use atomistic molecular dynamics
(MD) simulations to study the unbinding process. However, most
successful drugs unbind at timescales much longer than
milliseconds.\cite{copeland2006drug,copeland2015drug} Even with the
fastest available supercomputers, this makes it virtually impossible
to use MD simulations to obtain statistically reliable insight into
unbinding dynamics.

This timescale limitation makes it crucial to complement MD with
enhanced sampling techniques. These techniques accelerate the movement
between metastable states separated by high ($\gg k_B T$) barriers,
but still allow recovering the unbiased thermodynamics and kinetics.
While in principle one could construct Markov State Models (MSM)
\cite{msm_bowman} to study the unbinding dynamics from multiple short,
unbiased simulations without any enhanced sampling, the associated
high barriers typical for unbinding make this extremely difficult. As
such, reported applications of MSM to such problems have been
indirect, and instead of directly studying unbinding, these
studies\cite{trypsin_msm,plattner2015protein} have actually looked at
the drug binding problem where the barriers tend to be smaller.  To
directly simulate the unbinding process, it thus becomes unavoidable
to use enhanced sampling methods. \cite{pan_kinetics}

On the other hand, the use of enhanced sampling methods to study high
barrier systems has its own caveats. Many such methods involve
controlling the probability distribution along a low-dimensional
reaction coordinate (RC), which best captures all the relevant slow
degrees of freedom. Typically many such order parameters or
collective variables (CVs) are available that can distinguish between
various metastable states of the system at hand. For ligand unbinding
these CVs could include ligand-host relative displacement, their
conformations and their hydration states.  However, often the
fluctuations in these CVs can be coupled in a non-trivial manner, and
it can be tricky to select a RC without having a prescience of the CVs
whose fluctuations matter the most for driving the process of
interest.

In this work we aim to answer the following question: given a certain
choice of order parameters (or collective variables) for a ligand-host
system, what is the optimal 1-dimensional RC for unbinding that
  can be expressed as a linear combination of these collective
variables? We are especially interested in determining how wet
this RC is. Wetness here denotes the weight ascribed to the
  descriptor of the solvation state of the binding site,
relative to other descriptors contributing to the RC.  
This will indicate how important biasing
  water density fluctuations in the host binding pocket is to the
  kinetics of ligand unbinding. While it is well-known through various
theoretical, simulation and experimental studies that collective water
motion into/out of binding pockets is correlated with
unbinding/binding respectively \cite{pan_kinetics,mccammon_review,
setny2013solvent,mondal_fuller,mondal_friesner_jctc,fullerene,weiss2016solvent}, 
we wish to have a quantitative measure of the utility of biasing these water fluctuations 
  in the  sampling of ligand unbinding.

\begin{figure}
        \includegraphics[height=3.5in]{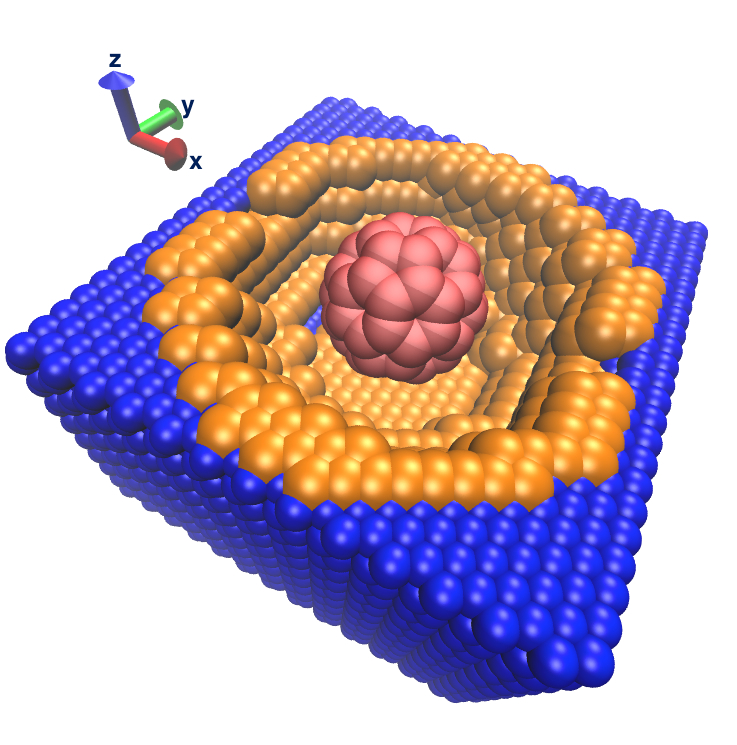} 
\caption { Cavity-ligand system in explicit water with axes
  marked. Red: fullerene shaped ligand atoms. Orange: cavity atoms
  that interact with the ligand and with water molecules. Blue: wall
  atoms. The water molecules are not shown for clarity. See
  Supplemental Information (SI) for corresponding interaction
  potentials.}
\label{system}
\end{figure}

Here, we investigate this question for ligand unbinding in a much
  studied model hydrophobic ligand-host system (Fig. \ref{system}) interacting through
  Lennard-Jones potential in an aqueous environment made of explicit
  TIP4P water molecules.\cite{fullerene,mondal_fuller,tip4p} Many excellent methods exist for
the purpose of RC optimization
\cite{besthummer_rc,coifman2005,peters_rc,ma_dinner,diffusionmap,noe_jcp_2013,
sketchmap,tuckerman_pnas2015,hummerszabo_dimred}. 
However, the energy barrier for unbinding in this system as reported through
previous studies is as high as 30 -- 35 $k_B T$, making it crucial for
the purpose of RC optimization to use a method that does not rely on
accurate sampling of rare reactive unbinding trajectories. For this
reason we use a recently proposed method SGOOP (Spectral gap
optimization of order parameters) \cite{sgoop} that enables us to
determine an optimal RC through relatively short biased
simulations performed using a trial RC (see Fig. \ref{flowchart} and Sec. \ref{sgoop} for
details of SGOOP).

We consider two different scenarios in this work, both of which are
expected to arise in the context of ligand unbinding. In the first
scenario, we sterically constrain the system so that the ligand can
move only along the centro-symmetric axis $z$ (see
Fig. \ref{system}). In the second, we lift this steric constraint. We
find that in the presence of the steric constraint, water density
fluctuations in the host cavity must be part of the optimal RC. This is
in excellent agreement with previous work on this and related systems
\cite{morrone2012interplay,li2012hydrodynamic,bolhuis_tps,
mondal_fuller,fullerene,patel2010fluctuations,gardehummerprl}
where for a sterically constrained set-up, there is a bimodal water
distribution at a critical ligand-cavity separation, around which the
unbinding pathway involves moving from dry to wet states. However we
find that when the steric constraint is removed and the
ligand is free to move in any direction, the role of water in the
optimal RC is minimal to none. In this case water is less of a driving
variable for unbinding, but more of a driven variable that follows the
movement of the ligand. Here SGOOP identifies how the optimal  RC is distorted  from
the $z-$axis (Fig.  \ref{system}), which
turns out to be the minimum free energy pathway for this system
as reported in a previous work.\cite{fullerene}

We validate our results through extensive calculations of unbinding
time statistics for the sterically constrained ligand using the
infrequent metadynamics approach \cite{meta_time} and find that the
optimal RC is indeed wet to some extent. In addition, because the
analogous barrier for ligand binding is much smaller than for
unbinding, we use unbiased MD estimates of the binding time and
validate that detailed balance is satisfied between unbinding and
binding rates. We perform the infrequent metadynamics calculations
using the optimal RC as per SGOOP and two other sub-optimal RCs with
no water content and more than optimal water content respectively. Our
findings clearly demonstrate the improvement in the quality and
accuracy of the unbinding time statistics by using the optimized RC
predicted through SGOOP. With the optimized CV the unbinding time
statistics gives a superior agreement with the binding time statistics
obtained through unbiased MD. Furthermore, it also gives a much
improved Poisson fit for the cumulative distribution function of
unbinding times, as quantified through the Kolmogorov-Smirnov test
proposed in Ref. \onlinecite{pvalue}. This shows that the optimized RC
predicted through SGOOP indeed does a better job of capturing the slow
dynamics of the system. Previous applications of SGOOP \cite{sgoop}
were restricted to using the optimized RC for faster convergence of
the free energy. The results reported in this work comprise the first
demonstration of improving kinetics calculations using SGOOP, and mark
a step further towards systematic high-throughput studies of unbinding
dynamics.
\section{Theory}
In this section we summarize the key methods \cite{sgoop,meta_time,pvalue}
used in this work and their underlying principles.

\subsection{Spectral gap optimization of order parameters (SGOOP)}
\label{sgoop}
\begin{figure}
\includegraphics[height=4.6in]{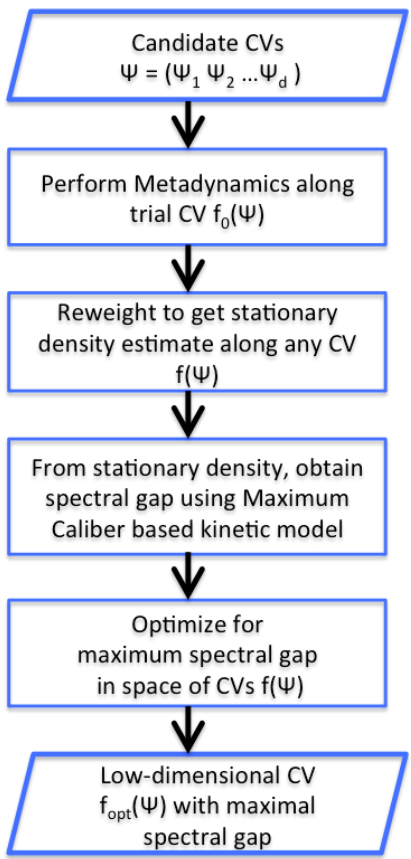} 
\caption {Flowchart summarizing the various key steps in
    SGOOP.\cite{sgoop} The whole process can in principle be iterated
    between the second and the last steps to further improve the
    sampling.}
\label{flowchart}
\end{figure}

SGOOP\cite{sgoop} is a method to optimize low-dimensional order
parameters or collective variables for use in enhanced sampling
biasing methods like umbrella sampling and metadynamics, when only
limited prior information is known about the system (see
  Fig. \ref{flowchart} for a flowchart summarizing the key steps in
  SGOOP). This optimization is done from a much larger set of
candidate CVs $\Psi = (\Psi_1,\Psi_2,...,\Psi_d)$, which are
assumed to be known \textit{a priori}. SGOOP is based on the idea that
the best order parameter, which we call the reaction coordinate (RC),
is one with the maximum separation of timescales between visible slow
and hidden fast processes. This timescale separation is calculated as
the spectral gap between the slow and fast eigenvalues of the
transition probability matrix on a grid along any CV \cite{sgoop}. The
transition probability matrix is calculated in SGOOP using an
approximate kinetic model that can be derived for example through the
principle of Maximum Caliber. \cite{sgoop,caliber1,dixit2015inferring}
Let $\{\lambda\}$ denote this set of eigenvalues, with $\lambda_0
\equiv 1 > \lambda_1 \geq \lambda_2 ...$. The spectral gap is then
defined as $\lambda_s - \lambda_{s+1} $, where $s$ is the number of
barriers apparent from the free energy estimate projected on the CV at
hand, that are higher than a user-defined threshold (typically
$\gtrsim k_B T$). In this case, assuming overdamped dynamics, the
eigenvalues beyond the first $s+1$ correspond to relaxation times in
each of the individual wells \cite{coifman,schuss,risken}, which for
an optimal RC should be much smaller than the escape times from the
wells.

The key input to SGOOP as used in this work is an estimate of the
stationary probability density (or equivalently the free energy) of
the system, accumulated through a biased simulation performed along a
sub-optimal trial RC given by some linear or non-linear function
$f_0(\Psi)$, where $\Psi$ denotes the larger set of candidate CVs.
Any type of biased simulation could be used for this purpose, as long
as it allows projecting the stationary probability density estimate on
generic combinations of CVs without having to repeat the
simulation. Metadynamics \cite{tiwary_rewt} provides this
functionality in a straightforward manner and hence we use it here.
Given this information we use the principle of Maximum Caliber
\cite{sgoop} to set up an unbiased master equation for the dynamics of
various trial CVs $f(\Psi)$. Through a post-processing optimization
procedure we then find the optimal RC as the $f(\Psi)$ which gives the
maximal spectral gap of the associated transfer matrix. We refer to
Ref. \onlinecite{sgoop} for details of the master equation and the
Maximum Caliber expression that relates the transfer matrix to
stationary probabilities, and facilitates calculation of the
eigenvalues and hence the spectral gap.

As described in the introduction, for the problem of ligand unbinding
in this work we take this larger set of CVs to be the various
components of the separation between the ligand and the host, and the
solvation state of the host pocket (Fig. \ref{system}). In more
complex systems, further members could be added to this set. Since
counting the number of barriers in a projected free energy profile
could be affected by sampling noise, we smooth the free energy by
averaging over bins. To ensure that the calculated spectral gaps are
robust with respect to amount of smoothening, we perform an averaged
estimate of the spectral gaps using different amounts of smoothing
(see SI for details).

Note that the approximate kinetic model used here in SGOOP is
equivalent to the Smoluchowski equation whereby (i) the dynamics of
any CV is described by a forced diffusion process, (ii) the diffusion
constant along this CV is independent of position. This kinetic model
is used in SGOOP to improve the choice of the RC that should be biased
given limited information starting with a trial RC.  The calculation
of rates is then done with this improved RC. It is important to note
that the infrequent metadynamics method for calculating rate constants
\cite{meta_time} \underline{does not} assume Smoluchowski dynamics or
constant diffusivity (see following Sec. \ref{metad} for details).

\begin{figure}
        \includegraphics[height=2.9in]{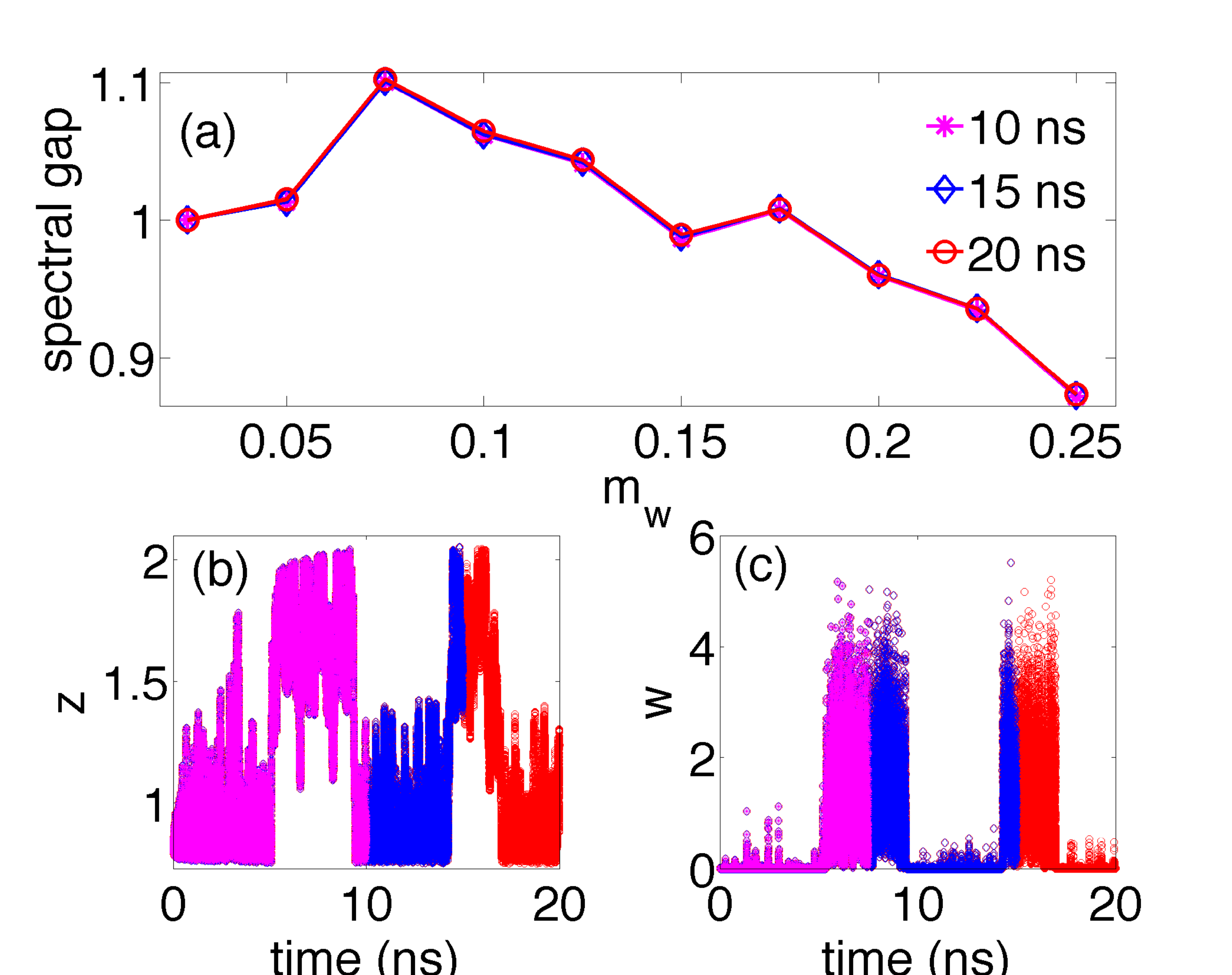} 
\caption { (a) Spectral gap versus amount of wetness of the RC, $m_w$
  (see Eq. \ref{cv_constr}) for the case when the ligand constrained
  to move along a line. The optimal RC can be clearly seen to be at
  $m_w \approx 0.075$. Three different profiles are provided, which
  were calculated by using starting metadynamics trajectories of
  different lengths as indicated in legends, performed with biasing CV
  $z$. The spectral gap is normalized so that its value for $m_w = 0$
  is 1.  (b)-(c) are the corresponding trajectories for the distance
  $z$ and number of pocket waters $w$. See Sec. \ref{results}A for
    precise definition of $w$. In all sub-figures here, red stars,
    blue diamonds and magenta circles denote results for trajectories
    of lengths 10 ns, 15 ns and 20 ns respectively.}
\label{sg_constr}
\end{figure}

\begin{figure*}
        \includegraphics[height=2.4in]{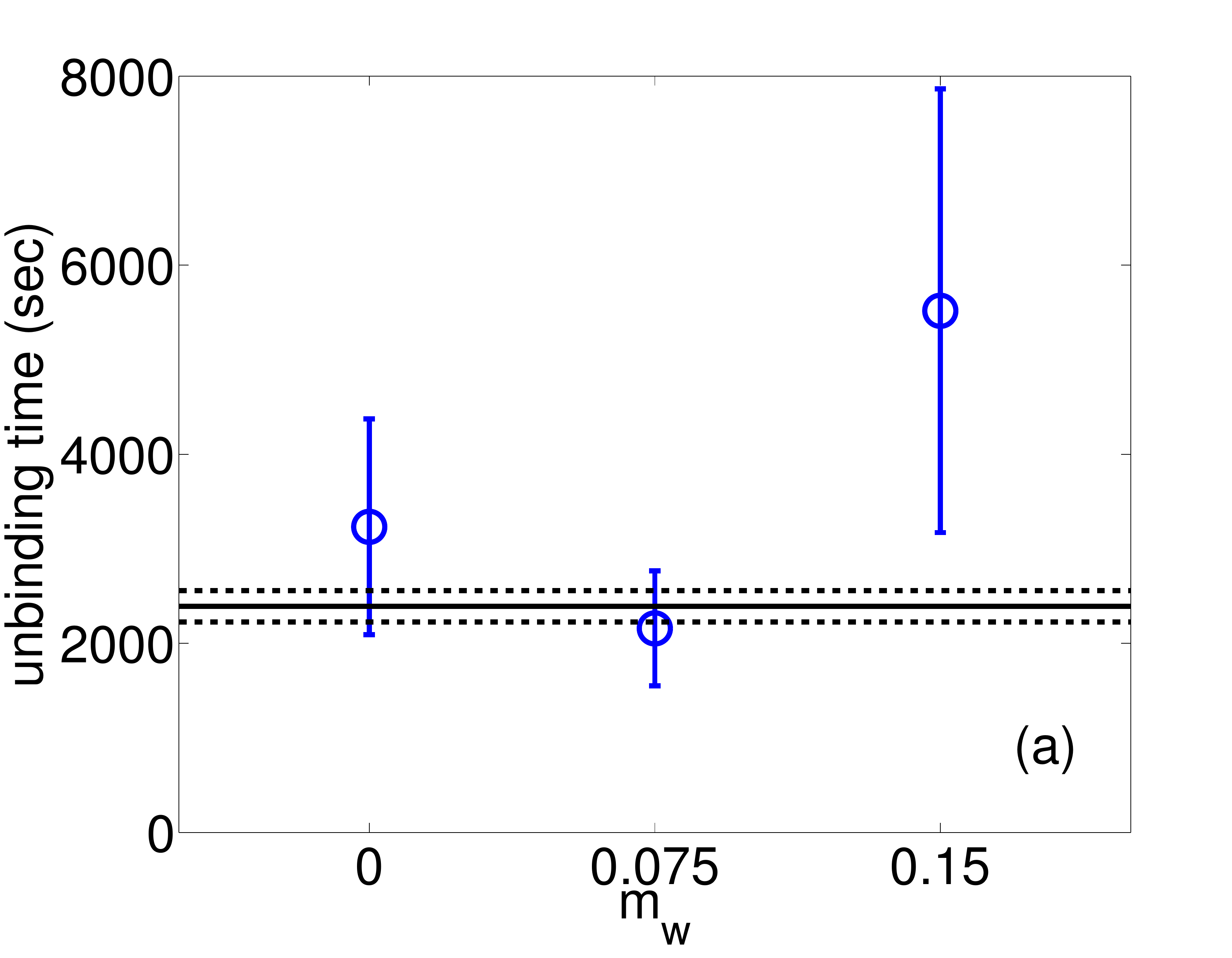} 
        \includegraphics[height=2.4in]{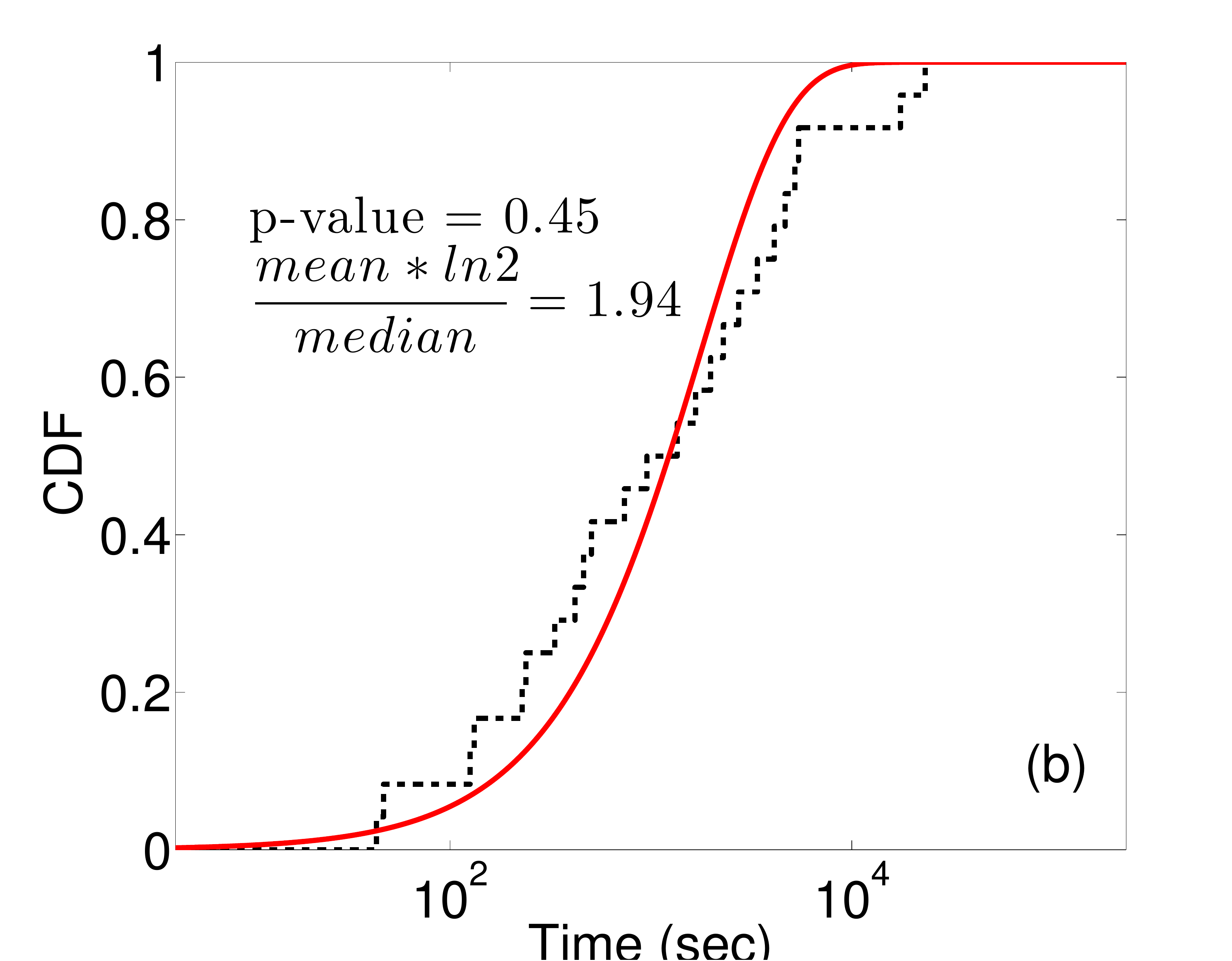} 
        \includegraphics[height=2.4in]{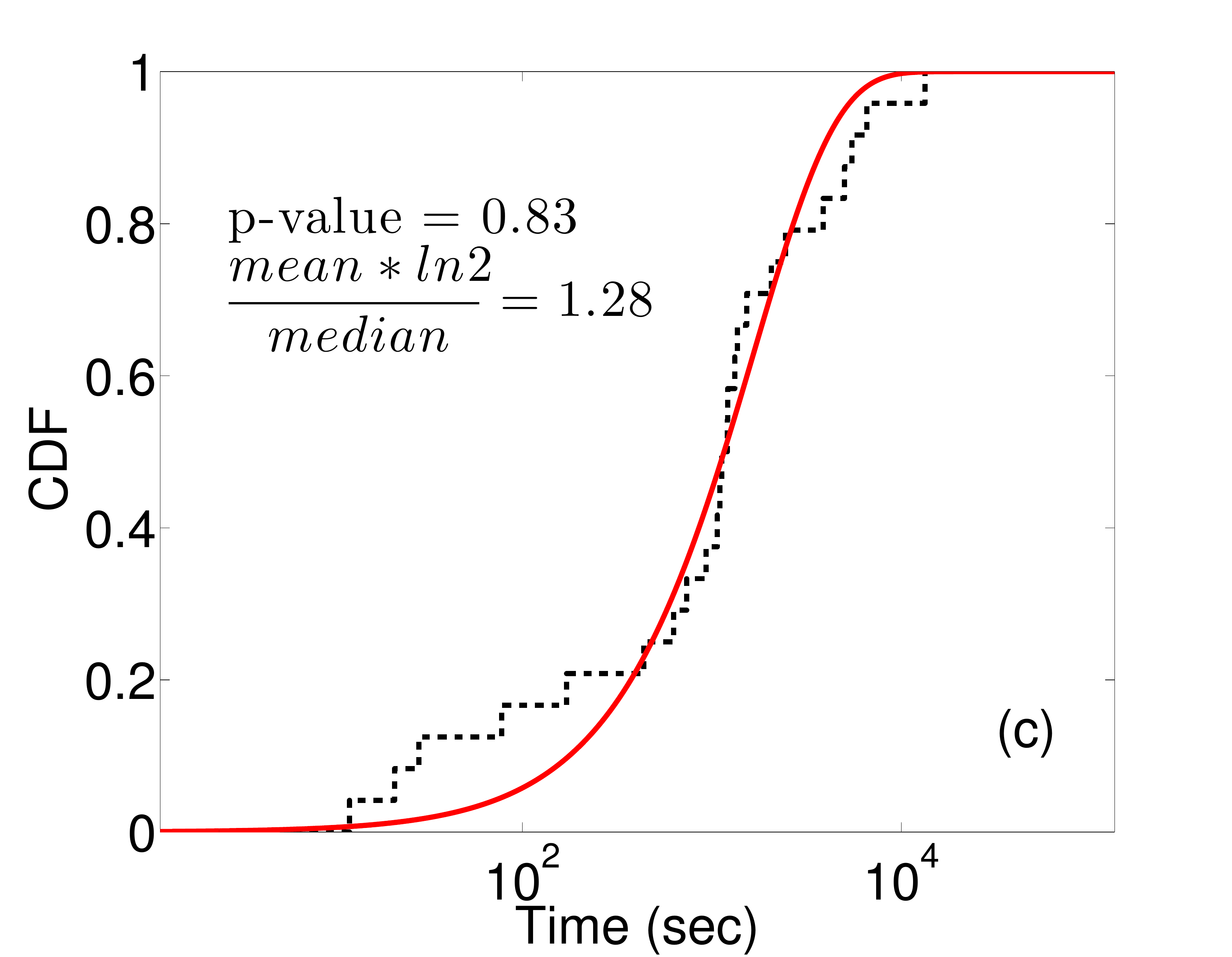} 
        \includegraphics[height=2.4in]{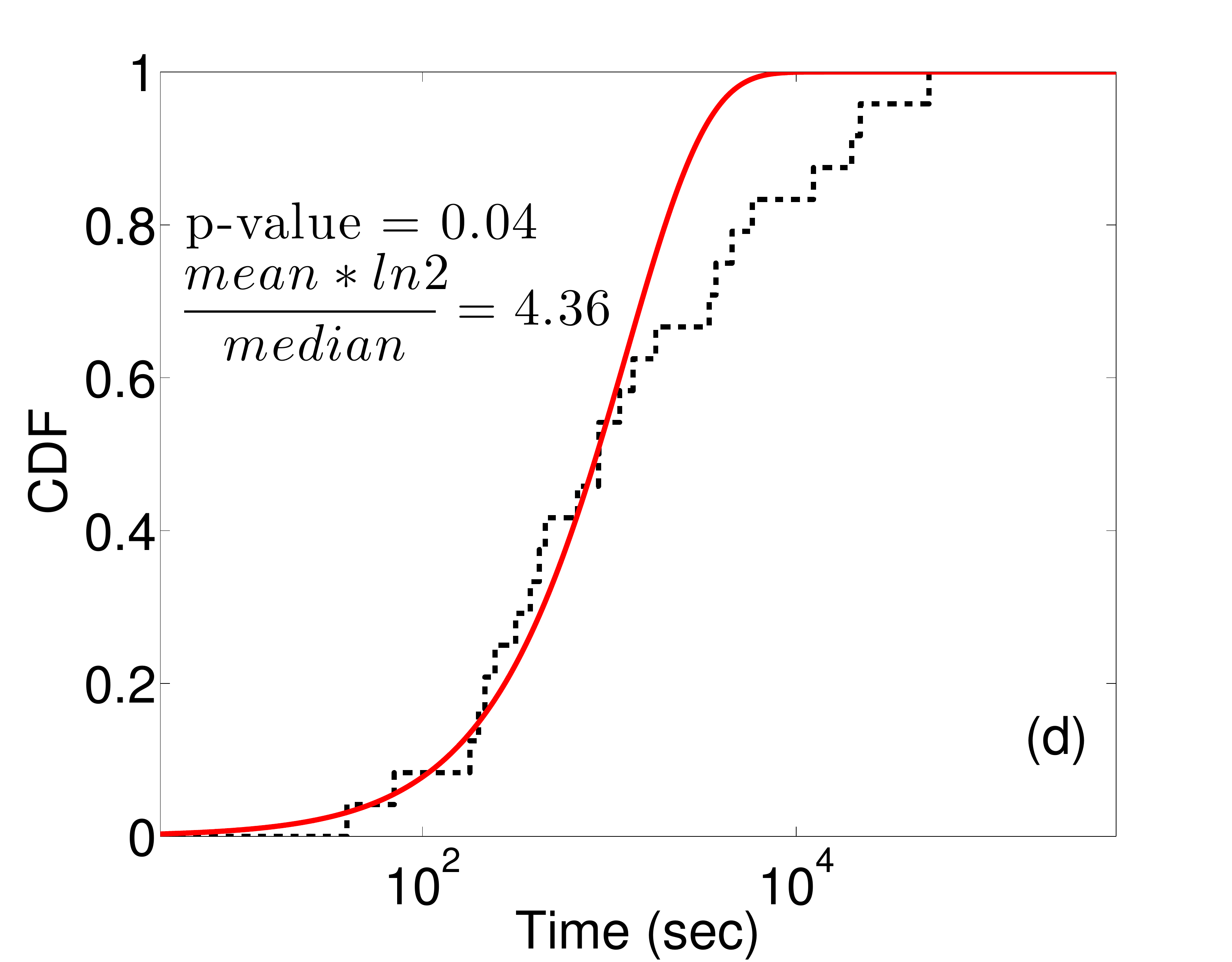} 
\caption { Unbinding times for the sterically constrained ligand using
  different simulation protocols. In (a) the mean unbinding times as
  obtained through the three RCs with different water coefficients are
  plotted along with error bars (blue circles). Also plotted is the
  corresponding estimate of mean unbinding time (solid black line)
  with errors (dashed black line) by using principle of detailed
  balance with the unbiased estimate of binding time. All error bars
  correspond to $\pm$ standard deviation intervals.  (b) to (d) give
  the empirical (black dashed line) and fitted (solid red line)
  cumulative distribution functions (CDF) for unbinding time
  statistics using different RCs with varying amounts of wetness $m_w$
  (see Eq. \ref{cv_constr}). From (b) to (d) respectively, $m_w$ is 0,
  0.075 and 0.15. Also indicated are respective p-values for fit to
  ideal Poisson distribution, quantified using the Kolmogorov-Smirnov
  test from Ref. \onlinecite{pvalue}, and mean times log(2) divided by
  median ratio for each case. The closer are both these values to 1,
  the more ideal is the Poisson distribution fit illustrating the
  reliability of the dynamics generated from metadynamics. As can be
  seen from these figures, the RC with optimal water coefficient of
  0.075 as obtained from SGOOP gives the best Poisson metric as per
  both these criteria.  }
\label{fits}
\end{figure*}

\subsection{Dynamics from infrequent metadynamics}
\label{metad}

The infrequent metadynamics approach\cite{meta_time,pvalue} is a
recently proposed method which has been used to obtain rate constants
in various molecular systems \cite{trypsin,fullerene}. It involves
time-dependent biasing of a few selected (typically one to three)
order parameters or collective variables (CVs) out of the many
available, in order to hasten the escape from metastable free
  energy basins. \cite{arpc_meta} By periodically adding repulsive
bias (typically in the form of Gaussians) in the regions of CV
space as they are visited, the system is encouraged to escape stable
free energy basins where they would normally be trapped for long
periods of time. The central idea in infrequent metadynamics
 is to deposit bias rarely enough
compared to the time spent in the transition state regions so that
dynamics in the saddle region is very rarely perturbed. Through this
approach one then increases the likelihood of not
corrupting the transition states, and preserves the sequence of
transitions between stable states. The acceleration of transition
rates achieved through biasing can then be calculated by appealing to
generalized transition state theory\cite{straubberne_review}, which
yields the following simple running average for the
acceleration\cite{meta_time}:
\begin{equation}
\alpha = \langle e^{\beta V(s,t)} \rangle _t
\label{acceleration}
\end{equation}
where $s$ is the collective variable being biased, $\beta = 1/ k_B T$
is the inverse temperature, $V(s,t)$ is the bias experienced at time
$t$ and the subscript $t$ indicates averaging under the time-dependent
potential. This approach is expected to work best in the diffusion
  controlled regime.\cite{kramers}

The infrequent metadynamics method requires a good and small set of slow collective variables
  demarcating all relevant stable states of interest. Whether this is
the case or not can be verified \textit{a posteriori} by checking if
the cumulative distribution function for the transition times out
  of each stable state is Poissonian\cite{pvalue}. While
metadynamics can still be performed with two, three, or more
biasing CVs, the computational gain obtained by compressing the slow
dynamics into an optimized 1-dimensional RC is immense, especially
given the infrequent nature of biasing (see SI for detailed simulation
parameters such as frequency of biasing used in this work).
Using SGOOP (Sec. \ref{sgoop}) allows us to select a good
  1-dimensional RC as a function of the many available choice of CVs,
  as we show in this work. This choice increases the probability of
  passing the test of Ref. \onlinecite{pvalue} once the relatively
  expensive infrequent metadynamics runs are performed.

\section{Results and discussion}
\label{results}
\subsection{Ligand constrained to move along one direction}
In the first case investigated, the system dynamics is sterically
  constrained so that the ligand can move only along the
  centro-symmetric axis $z$ (Fig. \ref{system}). This system and
  constraint has already been investigated in studies aimed at
  understanding hydrophobic interactions
  \cite{morrone2012interplay,li2012hydrodynamic,bolhuis_tps,mondal_fuller,fullerene}. Here
  we consider two descriptors; the $z$-component of the ligand-cavity
  separation, and the number of water molecules in the host cavity,
  denoted $w$. The number of water molecules is computed using a
  sigmoidal function which makes $w$ continuous and differentiable
  (see SI for details including precise definition of $w$) as
  implemented in the enhanced sampling plugin PLUMED \cite{plumed2}.
We then seek the best 1-d RC $f$ of the following form:
\begin{equation}
f(z,w) = \{ z + m_w w ; \text{           } m_w \geq 0 \}
\label{cv_constr}
\end{equation}
Throughout this paper $m_w$ is a measure of the wetness of the RC,
with $m_w=0$ corresponding to a completely dry RC, and higher values
denoting increasingly wetter RCs.

We first perform a short metadynamics simulation by biasing $f_0 =
z$. This starting run is performed with frequent biasing since the
objective here is to get a sense of the free energy, and not the
kinetics (see SI for various biasing frequencies and other parameters).
Through this we can obtain an estimate of the stationary
probability density along any $f(z,w)$ by using the reweighting
functionality of metadynamics \cite{tiwary_rewt}. By using SGOOP we
then get an estimate of the optimal $m_w \approx 0.075$ in
Eq. \ref{cv_constr} which maximizes the spectral gap. This is
  shown in Fig. \ref{sg_constr}(a) where an estimate of the spectral
  gap versus $m_w$ for different lengths of the starting metadynamics
  trajectory is provided. Other trajectories used in SGOOP shown in
  Fig. \ref{sg_constr} (b-c)) are for trajectories of length 10 ns, 15
  ns and 20 ns respectively. The results are extremely robust
  with respect to simulation time, and the spectral gaps estimated
  with trajectories of three different simulation times are
  virtually indistinguishable.

The optimal wetness of the RC in
Eq. \ref{cv_constr} given by $m_w \approx 0.075$ is validated by  by performing
extensive multiple independent unbinding simulations using infrequent
metadynamics, starting from the bound pose $z=0$ (Sec. \ref{metad}). The unbinding time
  is calculated as the time taken to reach $z=1.4$ nm for the first
  time.\cite{fullerene} We perform three independent sets of 24 simulations 
 (totaling 72 simulations) for:  (1) $m_w = 0$, a dry RC,
(2) $m_w = 0.075$, the RC with optimal  wetness found from SGOOP, and 
(3) $m_w = 0.15$, the RC with more than optimal
wetness.  The empirical and fitted cumulative distribution
functions for the unbinding time statistics using the three different
RCs with varying amounts of wetness are shown in  Figs. \ref{fits}
(b-d), along with the respective
p-values for fits to ideal Poisson distributions, quantified using the
Kolmogorov-Smirnov test from Ref. \onlinecite{pvalue}, and mean times
log(2) divided by median ratio for each case. An  ideal fit to the Poisson
distribution would result if  both these numbers would be close to 1, and this would 
suggest that the accelerated timescales found using  
metadynamics are reliable. The RC with optimal water coefficient $m_w$ = 0.075 obtained using
  SGOOP gives Poisson metrics closest to 1.  Fig. \ref{fits} (a)
shows the mean unbinding times obtained using the three RCs with
different values of the wetness parameter $m_w$ and these are compared
with the corresponding estimate provided in the
literature \cite{mondal_fuller} calculated from accurate free energy
 calculations together with the principle of
detailed balance. While it must be said that the completely dry
RC does a reasonable job in terms of the p-value and order of
magnitude agreement with unbiased MD, it is very clear from this plot
as well that the RC with optimal wetness gives the best performance as
per various metrics shown in Fig. \ref{fits}. Thus to summarize, the
optimal RC for this case indeed has a small but distinct amount of
wetness.

\begin{figure}
        \includegraphics[height=2.8in]{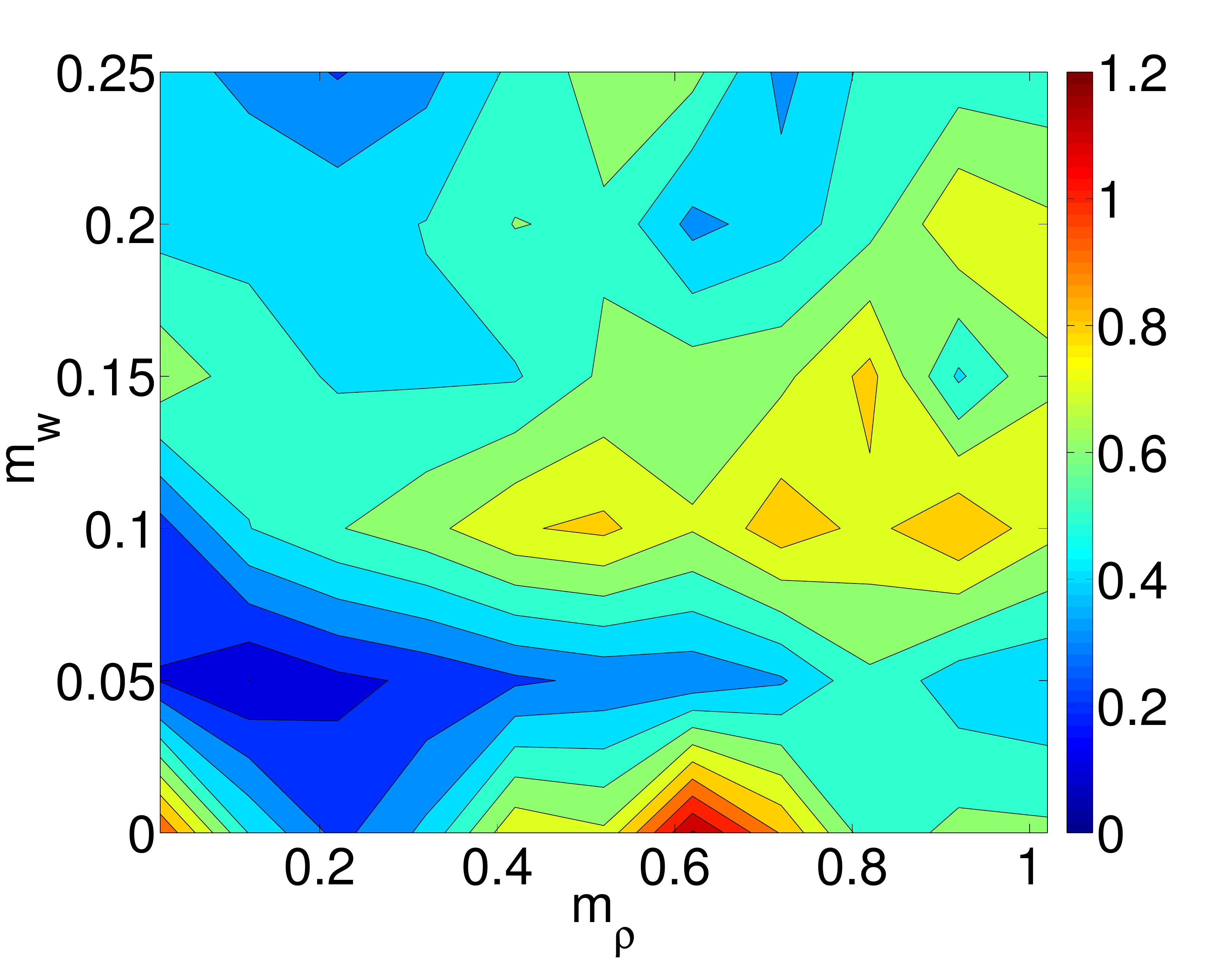} 
\caption { Contour plot of spectral gap versus $(m_{\rho},m_w)$ with
  starting metadynamics trajectory of duration 20 ns used in
  SGOOP. The optimal RC can be clearly seen to be at $(m_{\rho},m_w)
  \approx (0.6,0.0)$. The spectral gap is normalized so that its value
  for $(m_{\rho},m_w) = (0,0)$ is 1. }
\label{sg_noconstr}
\end{figure}

\subsection{Ligand free to move in any direction}
In this case case, we remove the steric constraint forcing the
  system to move along $z$, and allow the ligand to freely to move in
  any direction (see Fig. \ref{system}). Because the system is
axially symmetric, we consider 3 order parameters, namely the
$z$-component of the ligand-cavity separation, $\rho= \sqrt{x^2+y^2}$,
and the number of water molecules in the host cavity, denoted $w$. We
then seek the best 1-d RC $f$ of the following form:
\begin{equation}
f(z,\rho,w) = \{ z + m_{\rho} \rho + m_w w ; \text{           } m_{\rho} \geq 0, m_w \geq 0  \}
\label{cv_noconstr}
\end{equation}
We first perform a short metadynamics simulation by biasing with
  $f_0 = z$, a purely dry RC. As before, this starting run is
  performed with frequent biasing since the objective here is to get a
  sense of the free energy, and not the kinetics. This gives an
  estimate of the stationary probability density along any $f(z,\rho,w)$ by applying the reweighting functionality of
  metadynamics.\cite{tiwary_rewt} We then use SGOOP we to obtain an
  estimate of the optimal values as $ m_{\rho} \approx 0.6$, $m_w
  \approx 0.0$ in Eq. \ref{cv_noconstr}. These values maximize the
  spectral gap.

Fig. \ref{sg_noconstr} gives an estimate of the spectral gap
  versus $(m_{\rho},m_w)$ based on an initial metadynamics trajectory
  of duration 20 ns biasing $z$. The results are again
  extremely robust with respect to how long the simulation was
  run. See the SI for related data and other simulation
  parameters.

As can be seen by comparing Fig. \ref{sg_noconstr} to
  Fig. \ref{sg_constr}, the wetness of the the optimal RC in the case
  of unconstrained motion is closer to 0. In a sense the water
  fluctuations in the cavity appear to be caused or driven by the
  unbinding, rather than being a driving variable for unbinding as it is in
  the constrained case.  The primary reaction coordinate depends on
  $z$ and $\rho$, the displacement variables of the ligand with
  respect to the cavity. Indeed SGOOP finds $ m_{\rho} \approx 0.6$,
which gives the distortion of the reaction path from the $z-$axis (see
Fig.  \ref{system}). This is the same as the slope of the minimum free
energy pathway in $(z,\rho)$ space reported in previous work
\cite{fullerene}.

Since the optimal wetness of the RC in this case is close to 0, we do
not perform any kinetics calculations. Instead we refer to the results
from Ref. \onlinecite{fullerene}, where infrequent metadynamics with a
similar completely dry RC for this set-up gave very good agreement
through detailed balance with unbiased MD estimate of the binding
time.

\section{Discussion and Conclusions}
\label{conclusion}
In this work we have applied the recently proposed method SGOOP
\cite{sgoop} to the problem of determining the reaction coordinate for
ligand unbinding in a model system in explicit water. By using short
biased metadynamics simulations performed using a sub-optimal reaction
coordinate, we find that the true reaction coordinate involves water
in the case when the system is sterically constrained to move along an
axis of symmetry. In the case when this constraint is lifted, the role
of water in the optimal RC is reduced. Our predictions of the optimal
RC are validated by extensive calculations of the unbinding rate
constant using metadynamics with infrequent biasing
\cite{meta_time,pvalue} with different RCs. We believe that the
application of SGOOP to optimize the choice of RC for ligand
unbinding, combined with the approach of
Refs. \onlinecite{meta_time,pvalue}, provides an important step in the
quest to invent methods useful for systematic and possibly high
throughput calculations of the unbinding rate constant in more complex
and realistic protein-ligand systems, a quantity extremely difficult
to compute without careful enhanced sampling based approaches.
\cite{trypsin,trypsin_schulten} The hope is that this approach will
contribute a step toward the success of computational drug discovery
programs that take drug unbinding dynamics into account. We also think
 that the current work is a demonstration of
how SGOOP may be used to answer similar questions in systems other
than drug unbinding where the role of water density fluctuations in
driving the dynamics is believed to play a role but which is hard to
quantify.

 Using the model system in this work allows us to study an unbinding problem
involving solvation and steric related complexities, yet where we can
perform extensive simulations of the reverse binding
process. Undoubtedly more realistic systems will be harder to tackle
than the model system of the current work, possibly involving a much
larger set of trial collective variables than the current work, and
requiring more care in coming up with this trial set to begin with. As
long as the system's intrinsic dynamics displays a timescale
separation between few slow and remaining fast processes, and hence
possesses an associated spectral gap, we expect SGOOP to be useful in
obtaining a sense of fluctuations that matter for driving the dynamics
in rare event systems.

We would like to emphasize that the systems considered in this
  work, in spite of their model nature, are in fact quite challenging
  test cases. This is due to the enormous barrier height involved
  (around 30 -- 35 $k_B T$), and the relative insignificance of the
  barrier in the dewetting related bimodal distribution (around 1 -- 2
  $k_B T$)\cite{mondal_fuller} relative to this barrier. As such, even
  the trial RC that excludes wetness entirely, considered in this work
  and in Ref. \onlinecite{fullerene}, does a remarkably decent job
  when used with metadynamics.\cite{meta_time,pvalue} Yet SGOOP does
  very well in picking up signals in the right directions for
  improving the RC towards ideality. This demonstration makes us
  optimistic that in more complex systems where the barrier associated
  to movement of water is expected to be higher
  \cite{melittin,shaw_dasatinib,shaw_potassium,trypsin}, the algorithm
  will be even more useful. Some such studies are already underway and
  will be the subject of future publications.
\bigskip

\section*{ACKNOWLEDGMENTS}

This work was supported by grants from the National Institutes of
Health [NIH-GM4330] and the Extreme Science and Engineering Discovery
Environment (XSEDE) [TG-MCA08X002].

\bibliography{sgoop_fullerene}

\begin{thebibliography}{45}%
\makeatletter
\providecommand \@ifxundefined [1]{%
 \@ifx{#1\undefined}
}%
\providecommand \@ifnum [1]{%
 \ifnum #1\expandafter \@firstoftwo
 \else \expandafter \@secondoftwo
 \fi
}%
\providecommand \@ifx [1]{%
 \ifx #1\expandafter \@firstoftwo
 \else \expandafter \@secondoftwo
 \fi
}%
\providecommand \natexlab [1]{#1}%
\providecommand \enquote  [1]{``#1''}%
\providecommand \bibnamefont  [1]{#1}%
\providecommand \bibfnamefont [1]{#1}%
\providecommand \citenamefont [1]{#1}%
\providecommand \href@noop [0]{\@secondoftwo}%
\providecommand \href [0]{\begingroup \@sanitize@url \@href}%
\providecommand \@href[1]{\@@startlink{#1}\@@href}%
\providecommand \@@href[1]{\endgroup#1\@@endlink}%
\providecommand \@sanitize@url [0]{\catcode `\\12\catcode `\$12\catcode
  `\&12\catcode `\#12\catcode `\^12\catcode `\_12\catcode `\%12\relax}%
\providecommand \@@startlink[1]{}%
\providecommand \@@endlink[0]{}%
\providecommand \url  [0]{\begingroup\@sanitize@url \@url }%
\providecommand \@url [1]{\endgroup\@href {#1}{\urlprefix }}%
\providecommand \urlprefix  [0]{URL }%
\providecommand \Eprint [0]{\href }%
\providecommand \doibase [0]{http://dx.doi.org/}%
\providecommand \selectlanguage [0]{\@gobble}%
\providecommand \bibinfo  [0]{\@secondoftwo}%
\providecommand \bibfield  [0]{\@secondoftwo}%
\providecommand \translation [1]{[#1]}%
\providecommand \BibitemOpen [0]{}%
\providecommand \bibitemStop [0]{}%
\providecommand \bibitemNoStop [0]{.\EOS\space}%
\providecommand \EOS [0]{\spacefactor3000\relax}%
\providecommand \BibitemShut  [1]{\csname bibitem#1\endcsname}%
\let\auto@bib@innerbib\@empty
\bibitem [{\citenamefont {Copeland}, \citenamefont {Pompliano},\ and\
  \citenamefont {Meek}(2006)}]{copeland2006drug}%
  \BibitemOpen
  \bibfield  {author} {\bibinfo {author} {\bibfnamefont {R.~A.}\ \bibnamefont
  {Copeland}}, \bibinfo {author} {\bibfnamefont {D.~L.}\ \bibnamefont
  {Pompliano}}, \ and\ \bibinfo {author} {\bibfnamefont {T.~D.}\ \bibnamefont
  {Meek}},\ }\href@noop {} {\bibfield  {journal} {\bibinfo  {journal} {Nat.
  Rev. Drug. Discov.}\ }\textbf {\bibinfo {volume} {5}},\ \bibinfo {pages}
  {730} (\bibinfo {year} {2006})}\BibitemShut {NoStop}%
\bibitem [{\citenamefont {Copeland}(2015)}]{copeland2015drug}%
  \BibitemOpen
  \bibfield  {author} {\bibinfo {author} {\bibfnamefont {R.~A.}\ \bibnamefont
  {Copeland}},\ }\href@noop {} {\bibfield  {journal} {\bibinfo  {journal} {Nat.
  Rev. Drug. Discov.}\ } (\bibinfo {year} {2015})}\BibitemShut {NoStop}%
\bibitem [{\citenamefont {Bowman}\ \emph {et~al.}(2009)\citenamefont {Bowman},
  \citenamefont {Beauchamp}, \citenamefont {Boxer},\ and\ \citenamefont
  {Pande}}]{msm_bowman}%
  \BibitemOpen
  \bibfield  {author} {\bibinfo {author} {\bibfnamefont {G.~R.}\ \bibnamefont
  {Bowman}}, \bibinfo {author} {\bibfnamefont {K.~A.}\ \bibnamefont
  {Beauchamp}}, \bibinfo {author} {\bibfnamefont {G.}~\bibnamefont {Boxer}}, \
  and\ \bibinfo {author} {\bibfnamefont {V.~S.}\ \bibnamefont {Pande}},\
  }\href@noop {} {\bibfield  {journal} {\bibinfo  {journal} {J. Chem. Phys.}\
  }\textbf {\bibinfo {volume} {131}},\ \bibinfo {pages} {124101} (\bibinfo
  {year} {2009})}\BibitemShut {NoStop}%
\bibitem [{\citenamefont {Buch}, \citenamefont {Giorgino},\ and\ \citenamefont
  {De~Fabritiis}(2011)}]{trypsin_msm}%
  \BibitemOpen
  \bibfield  {author} {\bibinfo {author} {\bibfnamefont {I.}~\bibnamefont
  {Buch}}, \bibinfo {author} {\bibfnamefont {T.}~\bibnamefont {Giorgino}}, \
  and\ \bibinfo {author} {\bibfnamefont {G.}~\bibnamefont {De~Fabritiis}},\
  }\href@noop {} {\bibfield  {journal} {\bibinfo  {journal} {Proc. Natl. Acad.
  Sci.}\ }\textbf {\bibinfo {volume} {108}},\ \bibinfo {pages} {10184}
  (\bibinfo {year} {2011})}\BibitemShut {NoStop}%
\bibitem [{\citenamefont {Plattner}\ and\ \citenamefont
  {No{\'e}}(2015)}]{plattner2015protein}%
  \BibitemOpen
  \bibfield  {author} {\bibinfo {author} {\bibfnamefont {N.}~\bibnamefont
  {Plattner}}\ and\ \bibinfo {author} {\bibfnamefont {F.}~\bibnamefont
  {No{\'e}}},\ }\href@noop {} {\bibfield  {journal} {\bibinfo  {journal} {Nat.
  Comm.}\ }\textbf {\bibinfo {volume} {6}} (\bibinfo {year}
  {2015})}\BibitemShut {NoStop}%
\bibitem [{\citenamefont {Pan}\ \emph {et~al.}(2013)\citenamefont {Pan},
  \citenamefont {Borhani}, \citenamefont {Dror},\ and\ \citenamefont
  {Shaw}}]{pan_kinetics}%
  \BibitemOpen
  \bibfield  {author} {\bibinfo {author} {\bibfnamefont {A.~C.}\ \bibnamefont
  {Pan}}, \bibinfo {author} {\bibfnamefont {D.~W.}\ \bibnamefont {Borhani}},
  \bibinfo {author} {\bibfnamefont {R.~O.}\ \bibnamefont {Dror}}, \ and\
  \bibinfo {author} {\bibfnamefont {D.~E.}\ \bibnamefont {Shaw}},\ }\href@noop
  {} {\bibfield  {journal} {\bibinfo  {journal} {Drug Discov Today}\ }\textbf
  {\bibinfo {volume} {18}},\ \bibinfo {pages} {667} (\bibinfo {year}
  {2013})}\BibitemShut {NoStop}%
\bibitem [{\citenamefont {Baron}\ and\ \citenamefont
  {McCammon}(2013)}]{mccammon_review}%
  \BibitemOpen
  \bibfield  {author} {\bibinfo {author} {\bibfnamefont {R.}~\bibnamefont
  {Baron}}\ and\ \bibinfo {author} {\bibfnamefont {J.~A.}\ \bibnamefont
  {McCammon}},\ }\href@noop {} {\bibfield  {journal} {\bibinfo  {journal} {Ann.
  Rev. Phys. Chem.}\ }\textbf {\bibinfo {volume} {64}},\ \bibinfo {pages} {151}
  (\bibinfo {year} {2013})}\BibitemShut {NoStop}%
\bibitem [{\citenamefont {Setny}\ \emph {et~al.}(2013)\citenamefont {Setny},
  \citenamefont {Baron}, \citenamefont {Kekenes-Huskey}, \citenamefont
  {McCammon},\ and\ \citenamefont {Dzubiella}}]{setny2013solvent}%
  \BibitemOpen
  \bibfield  {author} {\bibinfo {author} {\bibfnamefont {P.}~\bibnamefont
  {Setny}}, \bibinfo {author} {\bibfnamefont {R.}~\bibnamefont {Baron}},
  \bibinfo {author} {\bibfnamefont {P.~M.}\ \bibnamefont {Kekenes-Huskey}},
  \bibinfo {author} {\bibfnamefont {J.~A.}\ \bibnamefont {McCammon}}, \ and\
  \bibinfo {author} {\bibfnamefont {J.}~\bibnamefont {Dzubiella}},\ }\href@noop
  {} {\bibfield  {journal} {\bibinfo  {journal} {Proc. Natl. Acad. Sci.}\
  }\textbf {\bibinfo {volume} {110}},\ \bibinfo {pages} {1197} (\bibinfo {year}
  {2013})}\BibitemShut {NoStop}%
\bibitem [{\citenamefont {Mondal}, \citenamefont {Morrone},\ and\ \citenamefont
  {Berne}(2013)}]{mondal_fuller}%
  \BibitemOpen
  \bibfield  {author} {\bibinfo {author} {\bibfnamefont {J.}~\bibnamefont
  {Mondal}}, \bibinfo {author} {\bibfnamefont {J.~A.}\ \bibnamefont {Morrone}},
  \ and\ \bibinfo {author} {\bibfnamefont {B.~J.}\ \bibnamefont {Berne}},\
  }\href@noop {} {\bibfield  {journal} {\bibinfo  {journal} {Proc. Natl. Acad.
  Sci.}\ }\textbf {\bibinfo {volume} {110}},\ \bibinfo {pages} {13277}
  (\bibinfo {year} {2013})}\BibitemShut {NoStop}%
\bibitem [{\citenamefont {Mondal}, \citenamefont {Friesner},\ and\
  \citenamefont {Berne}(2014)}]{mondal_friesner_jctc}%
  \BibitemOpen
  \bibfield  {author} {\bibinfo {author} {\bibfnamefont {J.}~\bibnamefont
  {Mondal}}, \bibinfo {author} {\bibfnamefont {R.~A.}\ \bibnamefont
  {Friesner}}, \ and\ \bibinfo {author} {\bibfnamefont {B.}~\bibnamefont
  {Berne}},\ }\href@noop {} {\bibfield  {journal} {\bibinfo  {journal} {J.
  Chem. Theor. Comp.}\ }\textbf {\bibinfo {volume} {10}},\ \bibinfo {pages}
  {5696} (\bibinfo {year} {2014})}\BibitemShut {NoStop}%
\bibitem [{\citenamefont {Tiwary}\ \emph
  {et~al.}(2015{\natexlab{a}})\citenamefont {Tiwary}, \citenamefont {Mondal},
  \citenamefont {Morrone},\ and\ \citenamefont {Berne}}]{fullerene}%
  \BibitemOpen
  \bibfield  {author} {\bibinfo {author} {\bibfnamefont {P.}~\bibnamefont
  {Tiwary}}, \bibinfo {author} {\bibfnamefont {J.}~\bibnamefont {Mondal}},
  \bibinfo {author} {\bibfnamefont {J.~A.}\ \bibnamefont {Morrone}}, \ and\
  \bibinfo {author} {\bibfnamefont {B.~J.}\ \bibnamefont {Berne}},\ }\href
  {\doibase 10.1073/pnas.1516652112} {\bibfield  {journal} {\bibinfo  {journal}
  {Proc. Natl. Acad. Sci.}\ } (\bibinfo {year} {2015}{\natexlab{a}}),\
  10.1073/pnas.1516652112}\BibitemShut {NoStop}%
\bibitem [{\citenamefont {Wei{\ss}}, \citenamefont {Setny},\ and\ \citenamefont
  {Dzubiella}(2016)}]{weiss2016solvent}%
  \BibitemOpen
  \bibfield  {author} {\bibinfo {author} {\bibfnamefont {R.~G.}\ \bibnamefont
  {Wei{\ss}}}, \bibinfo {author} {\bibfnamefont {P.}~\bibnamefont {Setny}}, \
  and\ \bibinfo {author} {\bibfnamefont {J.}~\bibnamefont {Dzubiella}},\
  }\href@noop {} {\bibfield  {journal} {\bibinfo  {journal} {J. Phys. Chem. B}\
  } (\bibinfo {year} {2016})}\BibitemShut {NoStop}%
\bibitem [{\citenamefont {Jorgensen}\ \emph {et~al.}(1983)\citenamefont
  {Jorgensen}, \citenamefont {Chandrasekhar}, \citenamefont {Madura},
  \citenamefont {Impey},\ and\ \citenamefont {Klein}}]{tip4p}%
  \BibitemOpen
  \bibfield  {author} {\bibinfo {author} {\bibfnamefont {W.~L.}\ \bibnamefont
  {Jorgensen}}, \bibinfo {author} {\bibfnamefont {J.}~\bibnamefont
  {Chandrasekhar}}, \bibinfo {author} {\bibfnamefont {J.~D.}\ \bibnamefont
  {Madura}}, \bibinfo {author} {\bibfnamefont {R.~W.}\ \bibnamefont {Impey}}, \
  and\ \bibinfo {author} {\bibfnamefont {M.~L.}\ \bibnamefont {Klein}},\
  }\href@noop {} {\bibfield  {journal} {\bibinfo  {journal} {J. Chem. Phys.}\
  }\textbf {\bibinfo {volume} {79}},\ \bibinfo {pages} {926} (\bibinfo {year}
  {1983})}\BibitemShut {NoStop}%
\bibitem [{\citenamefont {Best}\ and\ \citenamefont
  {Hummer}(2005)}]{besthummer_rc}%
  \BibitemOpen
  \bibfield  {author} {\bibinfo {author} {\bibfnamefont {R.~B.}\ \bibnamefont
  {Best}}\ and\ \bibinfo {author} {\bibfnamefont {G.}~\bibnamefont {Hummer}},\
  }\href@noop {} {\bibfield  {journal} {\bibinfo  {journal} {Proc. Natl. Acad.
  Sci.}\ }\textbf {\bibinfo {volume} {102}},\ \bibinfo {pages} {6732} (\bibinfo
  {year} {2005})}\BibitemShut {NoStop}%
\bibitem [{\citenamefont {Coifman}\ \emph {et~al.}(2005)\citenamefont
  {Coifman}, \citenamefont {Lafon}, \citenamefont {Lee}, \citenamefont
  {Maggioni}, \citenamefont {Nadler}, \citenamefont {Warner},\ and\
  \citenamefont {Zucker}}]{coifman2005}%
  \BibitemOpen
  \bibfield  {author} {\bibinfo {author} {\bibfnamefont {R.~R.}\ \bibnamefont
  {Coifman}}, \bibinfo {author} {\bibfnamefont {S.}~\bibnamefont {Lafon}},
  \bibinfo {author} {\bibfnamefont {A.~B.}\ \bibnamefont {Lee}}, \bibinfo
  {author} {\bibfnamefont {M.}~\bibnamefont {Maggioni}}, \bibinfo {author}
  {\bibfnamefont {B.}~\bibnamefont {Nadler}}, \bibinfo {author} {\bibfnamefont
  {F.}~\bibnamefont {Warner}}, \ and\ \bibinfo {author} {\bibfnamefont {S.~W.}\
  \bibnamefont {Zucker}},\ }\href@noop {} {\bibfield  {journal} {\bibinfo
  {journal} {Proc. Natl. Acad. Sci.}\ }\textbf {\bibinfo {volume} {102}},\
  \bibinfo {pages} {7426} (\bibinfo {year} {2005})}\BibitemShut {NoStop}%
\bibitem [{\citenamefont {Peters}\ and\ \citenamefont
  {Trout}(2006)}]{peters_rc}%
  \BibitemOpen
  \bibfield  {author} {\bibinfo {author} {\bibfnamefont {B.}~\bibnamefont
  {Peters}}\ and\ \bibinfo {author} {\bibfnamefont {B.~L.}\ \bibnamefont
  {Trout}},\ }\href@noop {} {\bibfield  {journal} {\bibinfo  {journal} {J.
  Chem. Phys.}\ }\textbf {\bibinfo {volume} {125}},\ \bibinfo {pages} {054108}
  (\bibinfo {year} {2006})}\BibitemShut {NoStop}%
\bibitem [{\citenamefont {Ma}\ and\ \citenamefont {Dinner}(2005)}]{ma_dinner}%
  \BibitemOpen
  \bibfield  {author} {\bibinfo {author} {\bibfnamefont {A.}~\bibnamefont
  {Ma}}\ and\ \bibinfo {author} {\bibfnamefont {A.~R.}\ \bibnamefont
  {Dinner}},\ }\href@noop {} {\bibfield  {journal} {\bibinfo  {journal} {The
  Journal of Physical Chemistry B}\ }\textbf {\bibinfo {volume} {109}},\
  \bibinfo {pages} {6769} (\bibinfo {year} {2005})}\BibitemShut {NoStop}%
\bibitem [{\citenamefont {Rohrdanz}\ \emph {et~al.}(2011)\citenamefont
  {Rohrdanz}, \citenamefont {Zheng}, \citenamefont {Maggioni},\ and\
  \citenamefont {Clementi}}]{diffusionmap}%
  \BibitemOpen
  \bibfield  {author} {\bibinfo {author} {\bibfnamefont {M.~A.}\ \bibnamefont
  {Rohrdanz}}, \bibinfo {author} {\bibfnamefont {W.}~\bibnamefont {Zheng}},
  \bibinfo {author} {\bibfnamefont {M.}~\bibnamefont {Maggioni}}, \ and\
  \bibinfo {author} {\bibfnamefont {C.}~\bibnamefont {Clementi}},\ }\href@noop
  {} {\bibfield  {journal} {\bibinfo  {journal} {J. Chem. Phys.}\ }\textbf
  {\bibinfo {volume} {134}},\ \bibinfo {pages} {124116} (\bibinfo {year}
  {2011})}\BibitemShut {NoStop}%
\bibitem [{\citenamefont {P{\'e}rez-Hern{\'a}ndez}\ \emph
  {et~al.}(2013)\citenamefont {P{\'e}rez-Hern{\'a}ndez}, \citenamefont {Paul},
  \citenamefont {Giorgino}, \citenamefont {De~Fabritiis},\ and\ \citenamefont
  {No{\'e}}}]{noe_jcp_2013}%
  \BibitemOpen
  \bibfield  {author} {\bibinfo {author} {\bibfnamefont {G.}~\bibnamefont
  {P{\'e}rez-Hern{\'a}ndez}}, \bibinfo {author} {\bibfnamefont
  {F.}~\bibnamefont {Paul}}, \bibinfo {author} {\bibfnamefont {T.}~\bibnamefont
  {Giorgino}}, \bibinfo {author} {\bibfnamefont {G.}~\bibnamefont
  {De~Fabritiis}}, \ and\ \bibinfo {author} {\bibfnamefont {F.}~\bibnamefont
  {No{\'e}}},\ }\href@noop {} {\bibfield  {journal} {\bibinfo  {journal} {J.
  Chem. Phys.}\ }\textbf {\bibinfo {volume} {139}},\ \bibinfo {pages} {015102}
  (\bibinfo {year} {2013})}\BibitemShut {NoStop}%
\bibitem [{\citenamefont {Ceriotti}, \citenamefont {Tribello},\ and\
  \citenamefont {Parrinello}(2011)}]{sketchmap}%
  \BibitemOpen
  \bibfield  {author} {\bibinfo {author} {\bibfnamefont {M.}~\bibnamefont
  {Ceriotti}}, \bibinfo {author} {\bibfnamefont {G.~A.}\ \bibnamefont
  {Tribello}}, \ and\ \bibinfo {author} {\bibfnamefont {M.}~\bibnamefont
  {Parrinello}},\ }\href@noop {} {\bibfield  {journal} {\bibinfo  {journal}
  {Proc. Natl. Acad. Sci.}\ }\textbf {\bibinfo {volume} {108}},\ \bibinfo
  {pages} {13023} (\bibinfo {year} {2011})}\BibitemShut {NoStop}%
\bibitem [{\citenamefont {Chen}, \citenamefont {Yu},\ and\ \citenamefont
  {Tuckerman}(2015)}]{tuckerman_pnas2015}%
  \BibitemOpen
  \bibfield  {author} {\bibinfo {author} {\bibfnamefont {M.}~\bibnamefont
  {Chen}}, \bibinfo {author} {\bibfnamefont {T.-Q.}\ \bibnamefont {Yu}}, \ and\
  \bibinfo {author} {\bibfnamefont {M.~E.}\ \bibnamefont {Tuckerman}},\
  }\href@noop {} {\bibfield  {journal} {\bibinfo  {journal} {Proc. Natl. Acad.
  Sci.}\ }\textbf {\bibinfo {volume} {112}},\ \bibinfo {pages} {3235} (\bibinfo
  {year} {2015})}\BibitemShut {NoStop}%
\bibitem [{\citenamefont {Hummer}\ and\ \citenamefont
  {Szabo}(2014)}]{hummerszabo_dimred}%
  \BibitemOpen
  \bibfield  {author} {\bibinfo {author} {\bibfnamefont {G.}~\bibnamefont
  {Hummer}}\ and\ \bibinfo {author} {\bibfnamefont {A.}~\bibnamefont {Szabo}},\
  }\href@noop {} {\bibfield  {journal} {\bibinfo  {journal} {The Journal of
  Physical Chemistry B}\ }\textbf {\bibinfo {volume} {119}},\ \bibinfo {pages}
  {9029} (\bibinfo {year} {2014})}\BibitemShut {NoStop}%
\bibitem [{\citenamefont {Tiwary}\ and\ \citenamefont
  {Berne}(2016{\natexlab{a}})}]{sgoop}%
  \BibitemOpen
  \bibfield  {author} {\bibinfo {author} {\bibfnamefont {P.}~\bibnamefont
  {Tiwary}}\ and\ \bibinfo {author} {\bibfnamefont {B.~J.}\ \bibnamefont
  {Berne}},\ }\href {\doibase 10.1073/pnas.1600917113} {\bibfield  {journal}
  {\bibinfo  {journal} {Proc. Natl. Acad. Sci.}\ }\textbf {\bibinfo {volume}
  {113}},\ \bibinfo {pages} {2839} (\bibinfo {year}
  {2016}{\natexlab{a}})}\BibitemShut {NoStop}%
\bibitem [{\citenamefont {Morrone}, \citenamefont {Li},\ and\ \citenamefont
  {Berne}(2012)}]{morrone2012interplay}%
  \BibitemOpen
  \bibfield  {author} {\bibinfo {author} {\bibfnamefont {J.~A.}\ \bibnamefont
  {Morrone}}, \bibinfo {author} {\bibfnamefont {J.}~\bibnamefont {Li}}, \ and\
  \bibinfo {author} {\bibfnamefont {B.~J.}\ \bibnamefont {Berne}},\ }\href@noop
  {} {\bibfield  {journal} {\bibinfo  {journal} {The Journal of Physical
  Chemistry B}\ }\textbf {\bibinfo {volume} {116}},\ \bibinfo {pages} {378}
  (\bibinfo {year} {2012})}\BibitemShut {NoStop}%
\bibitem [{\citenamefont {Li}, \citenamefont {Morrone},\ and\ \citenamefont
  {Berne}(2012)}]{li2012hydrodynamic}%
  \BibitemOpen
  \bibfield  {author} {\bibinfo {author} {\bibfnamefont {J.}~\bibnamefont
  {Li}}, \bibinfo {author} {\bibfnamefont {J.~A.}\ \bibnamefont {Morrone}}, \
  and\ \bibinfo {author} {\bibfnamefont {B.}~\bibnamefont {Berne}},\
  }\href@noop {} {\bibfield  {journal} {\bibinfo  {journal} {The Journal of
  Physical Chemistry B}\ }\textbf {\bibinfo {volume} {116}},\ \bibinfo {pages}
  {11537} (\bibinfo {year} {2012})}\BibitemShut {NoStop}%
\bibitem [{\citenamefont {Bolhuis}\ and\ \citenamefont
  {Chandler}(2000)}]{bolhuis_tps}%
  \BibitemOpen
  \bibfield  {author} {\bibinfo {author} {\bibfnamefont {P.~G.}\ \bibnamefont
  {Bolhuis}}\ and\ \bibinfo {author} {\bibfnamefont {D.}~\bibnamefont
  {Chandler}},\ }\href@noop {} {\bibfield  {journal} {\bibinfo  {journal} {The
  Journal of Chemical Physics}\ }\textbf {\bibinfo {volume} {113}},\ \bibinfo
  {pages} {8154} (\bibinfo {year} {2000})}\BibitemShut {NoStop}%
\bibitem [{\citenamefont {Patel}, \citenamefont {Varilly},\ and\ \citenamefont
  {Chandler}(2010)}]{patel2010fluctuations}%
  \BibitemOpen
  \bibfield  {author} {\bibinfo {author} {\bibfnamefont {A.~J.}\ \bibnamefont
  {Patel}}, \bibinfo {author} {\bibfnamefont {P.}~\bibnamefont {Varilly}}, \
  and\ \bibinfo {author} {\bibfnamefont {D.}~\bibnamefont {Chandler}},\
  }\href@noop {} {\bibfield  {journal} {\bibinfo  {journal} {J. Phys. Chem. B}\
  }\textbf {\bibinfo {volume} {114}},\ \bibinfo {pages} {1632} (\bibinfo {year}
  {2010})}\BibitemShut {NoStop}%
\bibitem [{\citenamefont {Hummer}\ and\ \citenamefont
  {Garde}(1998)}]{gardehummerprl}%
  \BibitemOpen
  \bibfield  {author} {\bibinfo {author} {\bibfnamefont {G.}~\bibnamefont
  {Hummer}}\ and\ \bibinfo {author} {\bibfnamefont {S.}~\bibnamefont {Garde}},\
  }\href@noop {} {\bibfield  {journal} {\bibinfo  {journal} {Physical review
  letters}\ }\textbf {\bibinfo {volume} {80}},\ \bibinfo {pages} {4193}
  (\bibinfo {year} {1998})}\BibitemShut {NoStop}%
\bibitem [{\citenamefont {Tiwary}\ and\ \citenamefont
  {Parrinello}(2013)}]{meta_time}%
  \BibitemOpen
  \bibfield  {author} {\bibinfo {author} {\bibfnamefont {P.}~\bibnamefont
  {Tiwary}}\ and\ \bibinfo {author} {\bibfnamefont {M.}~\bibnamefont
  {Parrinello}},\ }\href {\doibase 10.1103/PhysRevLett.111.230602} {\bibfield
  {journal} {\bibinfo  {journal} {Phys. Rev. Lett.}\ }\textbf {\bibinfo
  {volume} {111}},\ \bibinfo {pages} {230602} (\bibinfo {year}
  {2013})}\BibitemShut {NoStop}%
\bibitem [{\citenamefont {Salvalaglio}, \citenamefont {Tiwary},\ and\
  \citenamefont {Parrinello}(2014)}]{pvalue}%
  \BibitemOpen
  \bibfield  {author} {\bibinfo {author} {\bibfnamefont {M.}~\bibnamefont
  {Salvalaglio}}, \bibinfo {author} {\bibfnamefont {P.}~\bibnamefont {Tiwary}},
  \ and\ \bibinfo {author} {\bibfnamefont {M.}~\bibnamefont {Parrinello}},\
  }\href {\doibase 10.1021/ct500040r} {\bibfield  {journal} {\bibinfo
  {journal} {J. Chem. Theor. Comp.}\ }\textbf {\bibinfo {volume} {10}},\
  \bibinfo {pages} {1420} (\bibinfo {year} {2014})}\BibitemShut {NoStop}%
\bibitem [{\citenamefont {Press\'e}\ \emph {et~al.}(2013)\citenamefont
  {Press\'e}, \citenamefont {Ghosh}, \citenamefont {Lee},\ and\ \citenamefont
  {Dill}}]{caliber1}%
  \BibitemOpen
  \bibfield  {author} {\bibinfo {author} {\bibfnamefont {S.}~\bibnamefont
  {Press\'e}}, \bibinfo {author} {\bibfnamefont {K.}~\bibnamefont {Ghosh}},
  \bibinfo {author} {\bibfnamefont {J.}~\bibnamefont {Lee}}, \ and\ \bibinfo
  {author} {\bibfnamefont {K.~A.}\ \bibnamefont {Dill}},\ }\href {\doibase
  10.1103/RevModPhys.85.1115} {\bibfield  {journal} {\bibinfo  {journal} {Rev.
  Mod. Phys.}\ }\textbf {\bibinfo {volume} {85}},\ \bibinfo {pages} {1115}
  (\bibinfo {year} {2013})}\BibitemShut {NoStop}%
\bibitem [{\citenamefont {Dixit}\ \emph {et~al.}(2015)\citenamefont {Dixit},
  \citenamefont {Jain}, \citenamefont {Stock},\ and\ \citenamefont
  {Dill}}]{dixit2015inferring}%
  \BibitemOpen
  \bibfield  {author} {\bibinfo {author} {\bibfnamefont {P.~D.}\ \bibnamefont
  {Dixit}}, \bibinfo {author} {\bibfnamefont {A.}~\bibnamefont {Jain}},
  \bibinfo {author} {\bibfnamefont {G.}~\bibnamefont {Stock}}, \ and\ \bibinfo
  {author} {\bibfnamefont {K.~A.}\ \bibnamefont {Dill}},\ }\href@noop {}
  {\bibfield  {journal} {\bibinfo  {journal} {J. Chem. Theor. Comp.}\ }\textbf
  {\bibinfo {volume} {11}},\ \bibinfo {pages} {5464} (\bibinfo {year}
  {2015})}\BibitemShut {NoStop}%
\bibitem [{\citenamefont {Coifman}\ \emph {et~al.}(2008)\citenamefont
  {Coifman}, \citenamefont {Kevrekidis}, \citenamefont {Lafon}, \citenamefont
  {Maggioni},\ and\ \citenamefont {Nadler}}]{coifman}%
  \BibitemOpen
  \bibfield  {author} {\bibinfo {author} {\bibfnamefont {R.~R.}\ \bibnamefont
  {Coifman}}, \bibinfo {author} {\bibfnamefont {I.~G.}\ \bibnamefont
  {Kevrekidis}}, \bibinfo {author} {\bibfnamefont {S.}~\bibnamefont {Lafon}},
  \bibinfo {author} {\bibfnamefont {M.}~\bibnamefont {Maggioni}}, \ and\
  \bibinfo {author} {\bibfnamefont {B.}~\bibnamefont {Nadler}},\ }\href@noop {}
  {\bibfield  {journal} {\bibinfo  {journal} {Mult. Mod. Sim.}\ }\textbf
  {\bibinfo {volume} {7}},\ \bibinfo {pages} {842} (\bibinfo {year}
  {2008})}\BibitemShut {NoStop}%
\bibitem [{\citenamefont {Matkowsky}\ and\ \citenamefont
  {Schuss}(1981)}]{schuss}%
  \BibitemOpen
  \bibfield  {author} {\bibinfo {author} {\bibfnamefont {B.}~\bibnamefont
  {Matkowsky}}\ and\ \bibinfo {author} {\bibfnamefont {Z.}~\bibnamefont
  {Schuss}},\ }\href@noop {} {\bibfield  {journal} {\bibinfo  {journal} {SIAM
  Journal on Applied Mathematics}\ }\textbf {\bibinfo {volume} {40}},\ \bibinfo
  {pages} {242} (\bibinfo {year} {1981})}\BibitemShut {NoStop}%
\bibitem [{\citenamefont {Risken}(1984)}]{risken}%
  \BibitemOpen
  \bibfield  {author} {\bibinfo {author} {\bibfnamefont {H.}~\bibnamefont
  {Risken}},\ }\href@noop {} {\emph {\bibinfo {title} {Fokker-planck
  equation}}}\ (\bibinfo  {publisher} {Springer},\ \bibinfo {year}
  {1984})\BibitemShut {NoStop}%
\bibitem [{\citenamefont {Tiwary}\ and\ \citenamefont
  {Parrinello}(2014)}]{tiwary_rewt}%
  \BibitemOpen
  \bibfield  {author} {\bibinfo {author} {\bibfnamefont {P.}~\bibnamefont
  {Tiwary}}\ and\ \bibinfo {author} {\bibfnamefont {M.}~\bibnamefont
  {Parrinello}},\ }\href@noop {} {\bibfield  {journal} {\bibinfo  {journal} {J.
  Phys. Chem. B}\ }\textbf {\bibinfo {volume} {119}},\ \bibinfo {pages} {736}
  (\bibinfo {year} {2014})}\BibitemShut {NoStop}%
\bibitem [{\citenamefont {Tiwary}\ \emph
  {et~al.}(2015{\natexlab{b}})\citenamefont {Tiwary}, \citenamefont
  {Limongelli}, \citenamefont {Salvalaglio},\ and\ \citenamefont
  {Parrinello}}]{trypsin}%
  \BibitemOpen
  \bibfield  {author} {\bibinfo {author} {\bibfnamefont {P.}~\bibnamefont
  {Tiwary}}, \bibinfo {author} {\bibfnamefont {V.}~\bibnamefont {Limongelli}},
  \bibinfo {author} {\bibfnamefont {M.}~\bibnamefont {Salvalaglio}}, \ and\
  \bibinfo {author} {\bibfnamefont {M.}~\bibnamefont {Parrinello}},\
  }\href@noop {} {\bibfield  {journal} {\bibinfo  {journal} {Proc. Natl. Acad.
  Sci.}\ }\textbf {\bibinfo {volume} {112}},\ \bibinfo {pages} {E386} (\bibinfo
  {year} {2015}{\natexlab{b}})}\BibitemShut {NoStop}%
\bibitem [{\citenamefont {Valsson}, \citenamefont {Tiwary},\ and\ \citenamefont
  {Parrinello}(2016)}]{arpc_meta}%
  \BibitemOpen
  \bibfield  {author} {\bibinfo {author} {\bibfnamefont {O.}~\bibnamefont
  {Valsson}}, \bibinfo {author} {\bibfnamefont {P.}~\bibnamefont {Tiwary}}, \
  and\ \bibinfo {author} {\bibfnamefont {M.}~\bibnamefont {Parrinello}},\
  }\href@noop {} {\bibfield  {journal} {\bibinfo  {journal} {Annual Review of
  Physical Chemistry}\ }\textbf {\bibinfo {volume} {67}} (\bibinfo {year}
  {2016})}\BibitemShut {NoStop}%
\bibitem [{\citenamefont {Berne}, \citenamefont {Borkovec},\ and\ \citenamefont
  {Straub}(1988)}]{straubberne_review}%
  \BibitemOpen
  \bibfield  {author} {\bibinfo {author} {\bibfnamefont {B.~J.}\ \bibnamefont
  {Berne}}, \bibinfo {author} {\bibfnamefont {M.}~\bibnamefont {Borkovec}}, \
  and\ \bibinfo {author} {\bibfnamefont {J.~E.}\ \bibnamefont {Straub}},\
  }\href@noop {} {\bibfield  {journal} {\bibinfo  {journal} {J. Phys. Chem.}\
  }\textbf {\bibinfo {volume} {92}},\ \bibinfo {pages} {3711} (\bibinfo {year}
  {1988})}\BibitemShut {NoStop}%
\bibitem [{\citenamefont {Tiwary}\ and\ \citenamefont
  {Berne}(2016{\natexlab{b}})}]{kramers}%
  \BibitemOpen
  \bibfield  {author} {\bibinfo {author} {\bibfnamefont {P.}~\bibnamefont
  {Tiwary}}\ and\ \bibinfo {author} {\bibfnamefont {B.~J.}\ \bibnamefont
  {Berne}},\ }\href@noop {} {\bibfield  {journal} {\bibinfo  {journal} {J.
  Chem. Phys.}\ }\textbf {\bibinfo {volume} {144}},\ \bibinfo {pages} {134103}
  (\bibinfo {year} {2016}{\natexlab{b}})}\BibitemShut {NoStop}%
\bibitem [{\citenamefont {Tribello}\ \emph {et~al.}(2014)\citenamefont
  {Tribello}, \citenamefont {Bonomi}, \citenamefont {Branduardi}, \citenamefont
  {Camilloni},\ and\ \citenamefont {Bussi}}]{plumed2}%
  \BibitemOpen
  \bibfield  {author} {\bibinfo {author} {\bibfnamefont {G.~A.}\ \bibnamefont
  {Tribello}}, \bibinfo {author} {\bibfnamefont {M.}~\bibnamefont {Bonomi}},
  \bibinfo {author} {\bibfnamefont {D.}~\bibnamefont {Branduardi}}, \bibinfo
  {author} {\bibfnamefont {C.}~\bibnamefont {Camilloni}}, \ and\ \bibinfo
  {author} {\bibfnamefont {G.}~\bibnamefont {Bussi}},\ }\href@noop {}
  {\bibfield  {journal} {\bibinfo  {journal} {Comp. Phys. Comm.}\ }\textbf
  {\bibinfo {volume} {185}},\ \bibinfo {pages} {604} (\bibinfo {year}
  {2014})}\BibitemShut {NoStop}%
\bibitem [{\citenamefont {Teo}\ \emph {et~al.}(2016)\citenamefont {Teo},
  \citenamefont {Mayne}, \citenamefont {Schulten},\ and\ \citenamefont
  {Leli{\`e}vre}}]{trypsin_schulten}%
  \BibitemOpen
  \bibfield  {author} {\bibinfo {author} {\bibfnamefont {I.}~\bibnamefont
  {Teo}}, \bibinfo {author} {\bibfnamefont {C.~G.}\ \bibnamefont {Mayne}},
  \bibinfo {author} {\bibfnamefont {K.}~\bibnamefont {Schulten}}, \ and\
  \bibinfo {author} {\bibfnamefont {T.}~\bibnamefont {Leli{\`e}vre}},\
  }\href@noop {} {\bibfield  {journal} {\bibinfo  {journal} {Journal of
  Chemical Theory and Computation}\ } (\bibinfo {year} {2016})}\BibitemShut
  {NoStop}%
\bibitem [{\citenamefont {Liu}\ \emph {et~al.}(2005)\citenamefont {Liu},
  \citenamefont {Huang}, \citenamefont {Zhou},\ and\ \citenamefont
  {Berne}}]{melittin}%
  \BibitemOpen
  \bibfield  {author} {\bibinfo {author} {\bibfnamefont {P.}~\bibnamefont
  {Liu}}, \bibinfo {author} {\bibfnamefont {X.}~\bibnamefont {Huang}}, \bibinfo
  {author} {\bibfnamefont {R.}~\bibnamefont {Zhou}}, \ and\ \bibinfo {author}
  {\bibfnamefont {B.~J.}\ \bibnamefont {Berne}},\ }\href@noop {} {\bibfield
  {journal} {\bibinfo  {journal} {Nature}\ }\textbf {\bibinfo {volume} {437}},\
  \bibinfo {pages} {159} (\bibinfo {year} {2005})}\BibitemShut {NoStop}%
\bibitem [{\citenamefont {Shan}\ \emph {et~al.}(2011)\citenamefont {Shan},
  \citenamefont {Kim}, \citenamefont {Eastwood}, \citenamefont {Dror},
  \citenamefont {Seeliger},\ and\ \citenamefont {Shaw}}]{shaw_dasatinib}%
  \BibitemOpen
  \bibfield  {author} {\bibinfo {author} {\bibfnamefont {Y.}~\bibnamefont
  {Shan}}, \bibinfo {author} {\bibfnamefont {E.~T.}\ \bibnamefont {Kim}},
  \bibinfo {author} {\bibfnamefont {M.~P.}\ \bibnamefont {Eastwood}}, \bibinfo
  {author} {\bibfnamefont {R.~O.}\ \bibnamefont {Dror}}, \bibinfo {author}
  {\bibfnamefont {M.~A.}\ \bibnamefont {Seeliger}}, \ and\ \bibinfo {author}
  {\bibfnamefont {D.~E.}\ \bibnamefont {Shaw}},\ }\href@noop {} {\bibfield
  {journal} {\bibinfo  {journal} {Journal of the American Chemical Society}\
  }\textbf {\bibinfo {volume} {133}},\ \bibinfo {pages} {9181} (\bibinfo {year}
  {2011})}\BibitemShut {NoStop}%
\bibitem [{\citenamefont {Jensen}\ \emph {et~al.}(2012)\citenamefont {Jensen},
  \citenamefont {Jogini}, \citenamefont {Borhani}, \citenamefont {Leffler},
  \citenamefont {Dror},\ and\ \citenamefont {Shaw}}]{shaw_potassium}%
  \BibitemOpen
  \bibfield  {author} {\bibinfo {author} {\bibfnamefont {M.~{\O}.}\
  \bibnamefont {Jensen}}, \bibinfo {author} {\bibfnamefont {V.}~\bibnamefont
  {Jogini}}, \bibinfo {author} {\bibfnamefont {D.~W.}\ \bibnamefont {Borhani}},
  \bibinfo {author} {\bibfnamefont {A.~E.}\ \bibnamefont {Leffler}}, \bibinfo
  {author} {\bibfnamefont {R.~O.}\ \bibnamefont {Dror}}, \ and\ \bibinfo
  {author} {\bibfnamefont {D.~E.}\ \bibnamefont {Shaw}},\ }\href@noop {}
  {\bibfield  {journal} {\bibinfo  {journal} {Science}\ }\textbf {\bibinfo
  {volume} {336}},\ \bibinfo {pages} {229} (\bibinfo {year}
  {2012})}\BibitemShut {NoStop}%
\end{thebibliography}%


%
\end{document}